\documentstyle[amssymb,stmaryrd,psfig]{mnv2}   

\title[Isolated Galaxies]
{The fundamental plane  of isolated early-type galaxies}

\author[F.M. Reda {\it et al.\/}]
{Fatma M. Reda,$^{1,2}$\thanks{E-mail: freda@astro.swin.edu.eu (FMR); 
dforbes@swin.edu.au (DAF); george.hau@durham.ac.uk (GKTH)}
Duncan A. Forbes$^{1\star}$ and  
George K.T. Hau$^{3\star}$\\
 $^1$Centre for Astrophysics \& Supercomputing, Swinburne University, 
Hawthorn, VIC 3122, Australia\\
 $^2$Astronomy Department, National Research Institute of Astronomy and 
Geophysics, Helwan, Cairo 11421, Egypt\\
 $^3$Department of Physics, South Road, Durham, DH1 3LE, U.K.
 }

\begin{document} 
\maketitle  

\begin{abstract}

Here we present new measurements of effective radii, surface brightnesses and 
internal velocity dispersions for 23 isolated early-type galaxies. 
The photometric properties are derived from new multi-colour imaging of 10 
galaxies, whereas the central kinematics for 7 galaxies are taken from
forthcoming work by Hau \& Forbes.
These are supplemented with data from the literature. We reproduce the 
colour-magnitude and Kormendy relations and strengthen the result of Paper I
that isolated galaxies follow the same photometric relations as galaxies in
high density environments. 
We also find that some isolated galaxies reveal fine structure indicative 
of a recent merger while others appear undisturbed. 
We examine the Fundamental Plane in both traditional $R_e$, $\mu_e$ and 
$\sigma$ space and also $\kappa$-space. Most isolated galaxies follow the same 
Fundamental Plane tilt and scatter for galaxies in high density environments. 
However, a few galaxies notably deviate from the plane in the sense of having 
smaller $M/L$ ratios. This can be understood in terms of their younger stellar 
populations, which are presumably induced by a gaseous merger.
Overall, isolated galaxies have similar properties to those in 
groups and clusters with a slight enhancement in the frequency of recent 
mergers/interactions.
\end{abstract}


\section{Introduction}

The observational studies of Djorgovski \& Davis (1987) and Dressler (1987) 
found that elliptical galaxies are confined to a tight plane
defined by:
$R_e \propto \sigma ^\alpha ~  <\mu_e>^\beta$ in the 3 dimensional space of 
central velocity dispersion ($\sigma$),  effective radius ($R_e$) and 
mean surface brightness ($<\mu_e>$) enclosed by $R_e$, where $\alpha$ and 
$\beta$ are coefficients.
This plane is known as the Fundamental Plane (FP), and it has a small 
intrinsic scatter in its edge-on projection suggesting a 
strong regularity in the process of elliptical galaxy formation 
(J{\o}rgensen, Franx \& Kj{\ae}rgaard 1993; JFK93).

Theoretically, the FP can be derived from the scalar form of the virial 
theorem as:
$R_e =K_s~ \sigma^2 <\mu_e>^{-1} (M/L)^{-1}$, where 
$K_s$ is a structure parameter which depends on the luminosity, kinematic 
and  density structure of a galaxy, and $(M/L)$ is the mass-to-light ratio
(Djorgovski, de Carvalho \& Han 1988).
The observed and theoretical forms of the FP are identical only if 
the term $K_s~(M/L)^{-1}$ is a power-law function of $\sigma$ and/or $<\mu_e>$.
The observed intrinsic scatter of the FP implies a deviation of the 
relation from the pure power-law form. These deviations reflect the 
effects of galaxy formation and evolutionary processes. 
Assuming that early-type galaxies are homologous, i.e. have similar 
kinematic, luminosity, and density distributions so that $K_s=$ constant, 
then the FP reflects the evolution of the $M/L$ ratio, i.e. their 
stellar population and dark matter content. However, a combination of 
stellar population and non-homology dependence of the 
FP tilt is also plausible (Trujillo, Burkert \& Bell 2004).

Various theoretical studies have attempted to reproduce the FP in terms of the 
merger and/or collapse history of galaxies, and its subsequent effects on 
star formation (e.g. Hjorth \& Madsen 1995; Capelato, 
de Carvalho \& Carlberg 1995; Levine 1997; Bekki 1998; Dantas et al. 2003; 
Nipoti, Londrillo \& Ciotti 2003).
In a hierarchical universe, galaxies in low-density regions are the last to 
collapse and should thus be younger than galaxies which collapsed early in 
the higher density regions (e.g. Kauffmann et al. 2004).

The effect of local environment on a galaxy's formation history and 
fundamental parameters has been the motivation for many studies.
Environmental effects have been seen in galaxy scaling relations. 
For example, in a comparison study of early-type galaxies in the field, groups 
and rich 
clusters, de Carvalho \& Djorgovski (1992) found that field galaxies showed 
more intrinsic scatter in their properties than those in clusters, especially 
when stellar population variables were included. 
In a study of 9000 early-type galaxies from the Sloan Digital 
Sky Survey, Bernardi et al. (2003b) found that the scatter from the FP 
correlates weakly with the galaxy local environment. These variations are 
in the sense that galaxies in dense regions are slightly fainter, or have 
higher velocity dispersions, than in less dense regions 
(Bernardi  et al. 2003a).
The results of both de Carvalho \& Djorgovski (1992) and Bernardi et al. 
(2003b) suggest a more extended formation epoch for galaxies in the field 
versus those in clusters. 
This is supported by studies that measure the luminosity-weighted age of 
galaxies in different environments (e.g. Terlevich \& Forbes 2002; Proctor 
et al. 2004). 
Interestingly, Evstigneeva, Reshetnikov \& Sotnikova (2002) have reported no 
significant difference in the tilt and scatter of the Fundamental Plane for 
strongly interacting early-type galaxies. 

Previous studies of the FP in different environments have not been extended 
to the very low densities of truly isolated galaxies.
In such extreme low-density environments we can eliminate the effect of 
many physical processes, such as tidal interactions, ram-pressure stripping, 
strangulation, high-speed galaxy-galaxy interactions and ongoing mergers, all 
of which may affect the evolution of galaxies in denser environments. 

Here, we further study the sample of 36 early-type isolated galaxies 
introduced in Reda et al. (2004; Paper I). We present imaging in the $B$ and 
$R$-bands for another 10 galaxies in the sample. These images are 
used to study the internal morphological structure of these 10 galaxies and to
obtain their magnitudes and colours.
The main aspect of this work however is to investigate the FP for isolated 
galaxies 
and compare it with that for galaxies in higher density environments. 
Our photometric parameters are supplemented by kinematic data from Hau \& 
Forbes (inpreparation, hereafter HF05) and from the literature.
Based on  the results of this study, we briefly discuss the 
implications for the formation of isolated early-type galaxies.

\section{Photometric parameters}

\subsection{Observations and data reduction}

In this paper we present imaging in the $B$ and $R$-bands for 
10 galaxies in our sample (the original sample of 36 isolated galaxies was 
introduced in Paper I). 
Images were obtained using the Wide Field Imager (WFI) on the ESO/MPG 2.2-m
telescope on 2001 August 7-10. 
Each galaxy had many exposures which were combined to give an average 
integration time as summarized in Table 1.
The median seeing conditions over the three nights 
were $1.3^{''}$ in $B$ and $1.1^{''}$ in the $R$-band. 
Observations of standard star fields from Landolt (1992) were obtained during 
the three nights.
All galaxy and standard star fields were reduced using the 
{\small ESOWFI} package within {\small IRAF}. 

\subsection{Magnitudes, surface brightnesses and effective radii}
To obtain the photometric parameters of these galaxies, we followed the same 
technique as in Paper I and corrected them for Galactic extinction using 
values from  Schlegel, Finkbeiner \& Davis (1998). 
The {\small QPHOT} task in {\small IRAF} was used to obtain aperture 
magnitudes of each galaxy and compute the sky level in an annulus of 1000 
pixels and width of 50 pixels. Then, the total magnitudes (Table 1) were 
derived by fitting a curve-of-growth to the galaxy aperture magnitudes. 
To obtain the absolute magnitudes ($M_B$) listed in Table 2, we used the
distances quoted in Paper I.

\begin{table}
\begin{center}
\renewcommand{\arraystretch}{1.0}
\begin{tabular}{lccc}
\multicolumn{4}{c}{\bf Table 1. \small Photometric measurements.}\\
\hline
Galaxy & Filter & Magnitude & Exp.time \\ 
  &  & (mag) &   (sec) \\
\hline 
NGC 6653      & $B$ & 12.8 $\pm$ 0.1 & 120 \\
              & $R$ & 11.4 $\pm$ 0.1 & 120 \\
NGC 6776      & $B$ & 12.5 $\pm$ 0.1 & 180 \\
	      & $R$ & 10.9 $\pm$ 0.1 & 300 \\
NGC 6799      & $B$ & 13.3 $\pm$ 0.1 & 120 \\
	      & $R$ & 11.8 $\pm$ 0.1 & 120 \\
NGC 6849      & $B$ & 12.9 $\pm$ 0.1 & 120 \\
	      & $R$ & 11.6 $\pm$ 0.1 & 120 \\
NGC 7796      & $B$ & 12.2 $\pm$ 0.1 & 120 \\
	      & $R$ & 10.6 $\pm$ 0.1 & 120 \\
MCG-01-03-018 & $B$ & 13.4 $\pm$ 0.1 & 180 \\
	      & $R$ & 12.0 $\pm$ 0.1 & 120 \\
ESO107-G004   & $B$ & 12.6 $\pm$ 0.2 & 120 \\
	      & $R$ & 11.2 $\pm$ 0.2 & 120 \\
ESO153-G003   & $B$ & 13.7 $\pm$ 0.1 & 120 \\
              & $R$ & 12.1 $\pm$ 0.1 & 120 \\
ESO194-G021   & $B$ & 13.3 $\pm$ 0.1 & 120 \\
              & $R$ & 11.6 $\pm$ 0.1 & 120 \\
ESO462-G015   & $B$ & 12.5 $\pm$ 0.2 & 120 \\
              & $R$ & 11.0 $\pm$ 0.1 & 120 \\
\hline	       
\end{tabular}
\end{center}
Notes: Measured magnitudes and total exposure time 
in $B$ and $R$-bands for the galaxies observed on the ESO/MPG 2.2-m telescope.
\end{table}  

Again following the same technique as in Paper I, the {\small ISOPHOTE} 
package in 
{\small IRAF} was used to fit a smooth elliptical model to the galaxy image. 
During the modelling process, the galaxy centre, position angle and 
ellipticity were allowed to vary. 
The surface brightness profile of the galaxies were fit to a de Vaucouleurs 
$R^{1/4}$ law, from which we obtained the effective radius ($R_e$) and the 
mean effective surface brightness ($<\mu_e>$). The error estimate for $R_e$ 
and $<\mu_e>$ was based on the variation during the fitting procedure.

\subsection{Other sources of the photometric parameters}
In Paper I, we obtained $R_e$ and $<\mu_e>$ for 8 galaxies of our sample. 
These observations were obtained 
in the $B$ and $R$ filters using the Wide Field Imager (WFI) on the 3.9-m 
Anglo-Australian Telescope (AAT). For details of the data reduction and the 
derived photometric parameters see Paper I.

Additional data are taken from the photometric catalogue by Prugniel 
\& Heraudeau (1998; PH98) which includes total $B$ magnitudes and mean 
effective  surface brightnesses $<\mu _e>_B$ for 21 galaxies of our total 
sample. Prugniel \& Heraudeau fit their data using a 
linear interpolation between the de Vaucouleurs ($R^{1/4}$) 
and exponential profiles.
We calculate $R_e$ for galaxies from PH98 using their $B$ and $<\mu _e>_B$ 
values as: $\log R_e = (<\mu _e>_B - B - 5.885)/5$.
To check any systematic differences between our measurements and PH98 for 
$R_e$ and $<\mu_e>_B$, we have computed the 
quantity $\log R_e -0.352 <\mu_e>_B$ for the 12 galaxies in common with our 
observations. This is the edge-on projection of the FP from JFK93.
Fig. 1 shows this quantity for the 12 galaxies in common. The inset in Fig.1
shows the distribution of the residuals of our measurements from the
one-to-one relation. Our measurements show a slight systematic offset of 0.04
with small scatter of $1\sigma =0.07$ which indicates good 
agreement.
The final photometric parameters used in the present study are summarized in 
Table 2.
  
\begin{figure}
\centerline{\psfig{figure=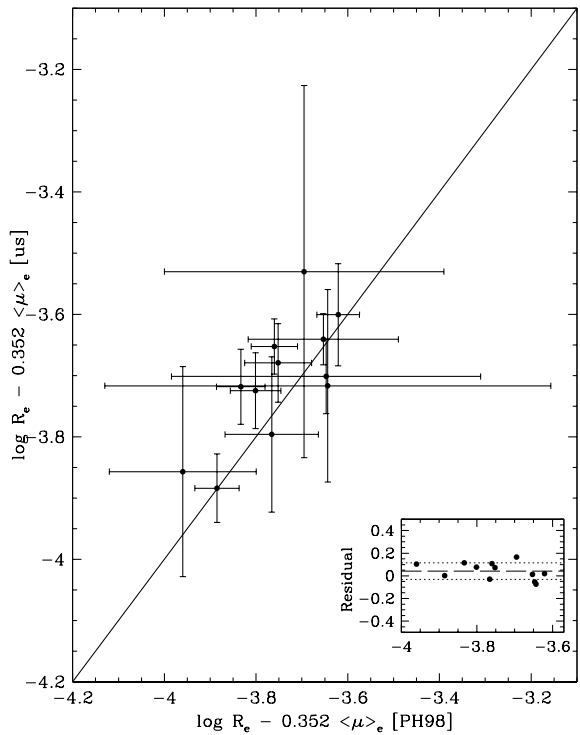,width=0.45\textwidth}}
\caption
{
Comparison of our measurements of $R_e$ and $<\mu_e>_B$ and those of Prugniel
\& Heraudeau (1998; PH98), using the edge-on projection of the Fundamental
Plane from J{\o}rgensen, Franx \& Kj{\ae}rgaard (1993). 
The solid line is the one-to-one relation. The inset shows the distribution of
the residuals of our measurements from the one-to-one relation. The 
long-dashed line is the mean value of the residuals and the dotted lines
represent the 1$\sigma$ dispersion. 
The common measurements are consistent within the errors.
}
\label{plot}
\end{figure}


\begin{figure}
\centerline{\psfig{figure=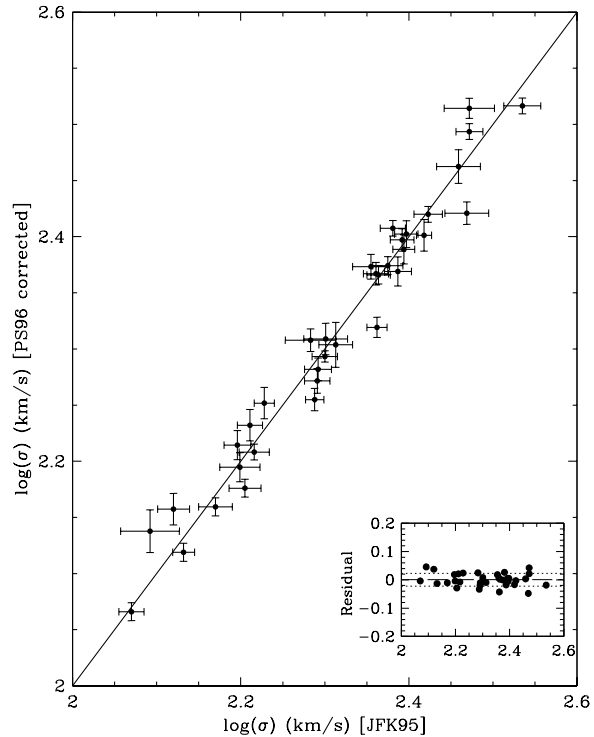,width=0.45\textwidth}}
\caption
{
Comparison between the velocity dispersion values of 35 galaxies from 
Prugniel \& Simien (1996; PS96) and values from J{\o}rgensen, Franx \& 
Kj{\ae}rgaard (1995; JFK95) after applying the transformation 
$\sigma_{JFK95}=-0.11+1.04 \times \sigma_{PS96}$. The solid line is the 
one-to-one relation. The inset shows the distribution of the residuals as in
Fig. 1. Good agreement is seen.
}
\label{plot}
\end{figure}

\begin{figure}
\centerline{\psfig{figure=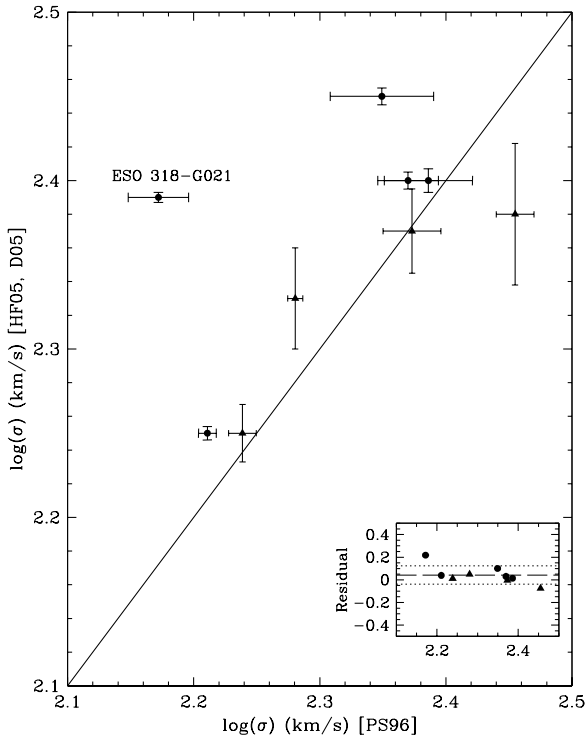,width=0.45\textwidth}}
\caption
{
Comparison between the velocity dispersion values from Hau \& Forbes 
(in preparation, hereafter HF05) shown as solid circles and Denicol\'{o} et al. (2005a; D05a) 
shown as solid triangles with those of Prugniel \& Simien (1996; PS96). 
For NGC 2271 we use the value quoted
in Koproline \& Zeilinger (2000).
The solid line is the one-to-one relation. The inset
shows the distribution of the residuals as in Fig. 1. Reasonable 
agreement is seen, except for ESO318-G021 (see text for
details). 
}
\label{plot}
\end{figure}

\section{Kinematic parameters}

\begin{table*}
\begin{center}
\renewcommand{\arraystretch}{1.0}
\begin{tabular}{lccccclccl}
\multicolumn{10}{c}{\bf Table 2. \small Photometric and spectroscopic properties of the sample.}\\
\hline
Galaxy & $M_B$ & $<\mu_e>_B$ & $\pm$ & $\log R_e$ & $\pm$ & Photometry & $\log (\sigma)$ &  $\pm$ & $\sigma$ source \\ 
 & (mag) & (mag/sq.$''$) &  & (pc) &   & source & (km/s) &  &   \\
\hline 
NGC 0821      & -20.5 & 21.6 &  0.03  & 3.70 & 0.03 & PH98    & 2.326 & 0.062 & D05a\\
NGC 1045      & -20.9 & 20.8 &  0.10  & 3.66 & 0.10 & Paper I & 2.360 & 0.061 & D05a\\
NGC 1132      & -21.8 & 22.5 &  0.10  & 4.12 & 0.05 & Paper I & 2.375 & 0.051 & D05a\\
NGC 2110      & -20.0 & 20.7 &  0.10  & 3.59 & 0.02 & Paper I & 2.400 & 0.007 & HF05\\
NGC 2271      & -20.0 & 20.9 &  0.06  & 3.47 & 0.07 & PH98    & 2.400 & 0.005 & HF05\\
NGC 2865      & -20.7 & 20.8 &  0.10  & 3.68 & 0.01 & Paper I & 2.250 & 0.004 & HF05\\
NGC 3562$^*$  & -21.8 & 21.8 &  0.09  & 4.02 & 0.07 & PH98    & 2.432 & 0.016 & PS96\\ 
NGC 6172      & -20.4 & 20.8 &  0.10  & 3.61 & 0.06 & Paper I & 2.130 & 0.038 & D05a\\
NGC 6411$^*$      & -21.2 & 21.9 &  0.12  & 3.90 & 0.04 & PH98    & 2.248 & 0.034 & D05a\\ 
NGC 6653      & -21.1 & 21.6 &  0.10  & 3.92 & 0.02 & ESO     & 2.337 & 0.041 & PS96\\
NGC 6702      & -21.5 & 22.3 &  0.05  & 4.04 & 0.04 & PH98    & 2.243 & 0.011 & PS96\\
NGC 6776      & -21.7 & 21.1 &  0.10  & 3.90 & 0.11 & ESO     & 2.316 & 0.028 & PS96\\
NGC 6799      & -21.1 & 21.3 &  0.10  & 3.77 & 0.02 & ESO     & 2.178 & 0.015 & PS96\\
NGC 6849      & -21.6 & 22.2 &  0.10  & 4.16 & 0.01 & ESO     & 2.328 & 0.014 & PS96\\
NGC 7796      & -21.0 & 21.1 &  0.10  & 3.71 & 0.02 & ESO     & 2.414 & 0.015 & PS96\\ 
IC 1211$^*$       & -21.0 & 21.9 &  0.03  & 3.88 & 0.06 & PH98    & 2.295 & 0.028 & PS96\\
MCG-01-27-013 & -21.4 & 22.0 &  0.10  & 3.98 & 0.05 & Paper I & 2.390 & 0.006 & HF05\\
MCG-01-03-018 & -21.0 & 22.3 &  0.10  & 3.96 & 0.01 & ESO     & 2.281 & 0.031 & D05a\\
MCG-03-26-030 & -21.7 & 21.4 &  0.10  & 3.92 & 0.02 & Paper I & 2.500 & 0.005 & HF05\\
ESO107-G004   & -20.4 & 21.4 &  0.20  & 3.68 & 0.06 & ESO     & 2.242 & 0.114 & PS96\\
ESO218-G002   & -20.9 & 21.5 &  0.10  & 3.76 & 0.11 & Paper I & 2.450 & 0.005 & HF05\\
ESO318-G021   & -20.7 & 21.7 &  0.04  & 3.77 & 0.01 & PH98    & 2.390 & 0.003 & HF05\\ 
ESO153-G003   & -20.9 & 20.6 &  0.10  & 3.62 & 0.06 & ESO     &   -   &   -   &  -  \\
ESO194-G021   & -19.7 & 21.0 &  0.10  & 3.47 & 0.02 & ESO     &   -   &   -   &  -  \\ 
ESO462-G015   & -22.0 & 21.3 &  0.20  & 3.90 & 0.02 & ESO     & 2.378 & 0.088 & D05a\\
\hline
\end{tabular} 
\end{center}
\begin {minipage} {150mm}
Notes: The adopted data for the Fundamental Plane study. The sources of the 
photometric parameters are Reda et al. (2004; Paper I), Prugniel \& 
Heraudeau (1998; PH98) and images obtained using the WFI on the ESO/MPG 
2.2-m telescope for galaxies listed in Table 1 (ESO). Sources for the 
velocity dispersion are Hau \& Forbes (2005; HF05), Denicol\'{o} et al. 
(2005a; D05a) and Prugniel \& Simien (1996; PS96). The two galaxies 
ESO153-G003 and ESO194-G021 do not have available velocity dispersions. 
The three galaxies marked with ''$^*$'' do not have $R$ magnitude in PH98 and 
are excluded from the colour-magnitude analysis.
\end {minipage}
\end{table*}

HF05 have conducted a detailed kinematical study for 9 
galaxies of our 36 isolated early-type galaxies. The galaxies were observed 
for $2 \times 1200$ seconds at the ESO 3.6-m telescope
on the La Silla Observatory, Chile
using a long slit of $1.5^{''}$ and Grism\#8. The slit was positioned roughly 
along the major axes of the galaxies and the spectra were summed inside a 
rectangular aperture of size $1.5^{''} \times 5^{''}$. The velocity dispersion 
($\sigma$) measurements were normalised to a diameter of $3.4^{''}$ for a 
galaxy at the distance of Coma using the method of J{\o}rgensen, Franx \& 
Kj{\ae}rgaard (1995; JFK95). For full details, see HF05.

In a recent study, Denicol\'{o} et al. (2005a; D05a) measured $\sigma$ 
for 8 early-type galaxies of our isolated sample. They used the same 
normalization technique of JFK95 as in HF05.
The galaxies were observed at the 2.12-m telescope of 
{\it The Observatorio Astrof$\acute{\iota}$sico Guillermo Haro}, 
in Cananea, Mexico. On average, they have eight observations for each galaxy 
in their study. 
These repeated observations per galaxy, and the deviation from the mean value,
give an error on the mean velocity dispersion quoted.

Another source of internal kinematics for our galaxy sample
is Prugniel \& Simien (1996; PS96). In this catalogue, Prugniel \& Simien 
compiled central velocity dispersions of 1698 galaxies from 3147 measurements. 
To homogenise the data that they collected from different literature sources, 
they identified a set of galaxies that have 
measurements from three or more references with deviations of 20 km/s or less. 
Using this list of ``standard'' galaxies, a scale factor was
determined for each source and used to scale the velocity dispersion 
measurements for that source. Then, $\sigma$ values were computed as the mean 
of the re-scaled measured velocity dispersion, weighted by the inverse of 
the squared mean error for each source.
Errors in the PS96 catalogue are 1$\sigma$ rms from the mean value.

The $\sigma$ values quoted in PS96 were not normalized to the aperture 
size as in HF05 and D05a. In JFK95, they applied the 
aperture normalization technique to 220 galaxies in a range of environments. 
Their sample has 35 galaxies in common (after excluding galaxies with low S/N 
ratio) with the sample of PS96. 
Using the $\sigma$ measurements of these common galaxies from 
JFK95 ($\sigma_{JFK95}$) and PS96 ($\sigma_{PS96}$), we obtained the
transformation: $\log\sigma_{JFK95}=-0.11+1.04 \times \log\sigma_{PS96}$, 
which is used to convert the $\sigma$ values from PS96 to be consistent with 
those of HF05 and D05a.
These 35 galaxies cover the redshift range from $cz \approx 750$ km/s to 
6500 km/s, which encompasses the range for our isolated galaxy sample. 
Any redshift-dependent aperture effect on the $\sigma$ values is less than 
the dispersion of $\pm$ 10 km/s about the one-to-one line shown in Fig. 2.
The inset in Fig.2 shows the distribution of the residuals of the corrected
PS96 measurements from the one-to-one relation. The mean value of the
residuals show a slight systematic offset of 0.0003 from the one-to-one 
relation with small scatter of $1\sigma =0.022$ which indicates good agreement.

The catalogue of PS96 contains velocity dispersions for 18 galaxies 
of our sample.
For the common galaxies between PS96 and HF05 (5 galaxies) and D05a 
(4 galaxies), we use the values from HF05 and D05a.
Fig. 3 shows a comparison between the $\sigma$ values from HF05 and D05a with
PS96 for the 8 galaxies in common. For the galaxy NGC 2271, we
use the velocity dispersion from Koproline \& Zeilinger (2000) who
quote a similar value to HF05. 

The most deviant galaxy in Fig. 3 is the galaxy ESO318-G021 which has a
velocity dispersion from PS96 that is significantly smaller than that from
HF05. We have examined the possible cause for this discrepancy and conclude
it is largely due to a mismatch of the template standard star. In HF05, a
best-fit to 
the spectrum of ESO318-G021 was achieved using a K4 giant star, however
when a K1 or K2 giant was used the velocity dispersion quoted by PS96 was
reproduced (albeit with a higher chi-squared value). The exclusion of
lines that are affected by emission increases the derived velocity
dispersion, but the main difference between PS96 and the HF05 is due to
the choice of stellar template. In the subsequent analysis we adopt the
HF05 velocity dispersion for ESO318-G021.
Considering all 9 galaxies in Fig. 3, the inset shows a mean value of
the residuals from the one-to-one line of 0.04 with a scatter of 1$\sigma
=0.08$.

\section{Results}

\subsection{Colour-magnitude relation}

\begin{figure}
\centerline{\psfig{figure=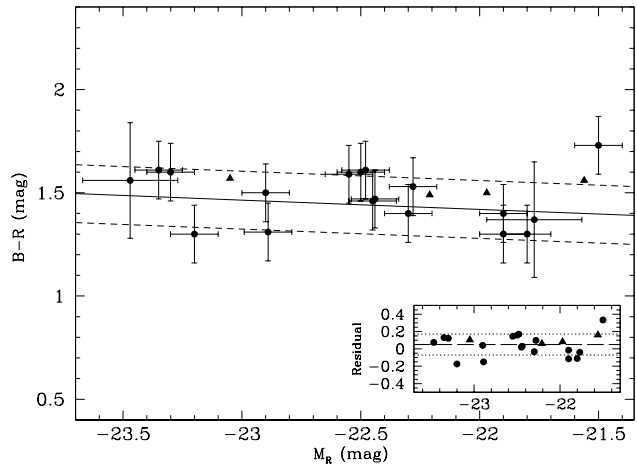,width=0.5\textwidth,angle=-90}}
\caption
{
Colour-magnitude relation for the sample galaxies. The solid line (Gladders et 
al. 1998) represents the CMR of elliptical galaxies in the Coma cluster, 
while the short-dashed line indicates the mean error in the colour
measurements for our sample galaxies. Solid circles are galaxies observed at
the AAT (Paper I) and ESO/MPG 2.2-m telescope (this work), while solid
triangles are galaxies from PH98 (error bars are not shown). The inset shows
the distribution of the residuals as in Fig. 1.
The isolated galaxies are consistent with the Coma cluster CMR. 
}
\label{plot}
\end{figure}

In Paper I, using the absolute magnitude in the $R$-band ($M_R$) and the 
($B-R$) colour of 8 isolated galaxies, we showed that they follow a 
colour-magnitude relation (CMR) of slope and intrinsic scatter similar to that 
of early-type galaxies in the denser environment of the Coma cluster. 
In Fig. 4 we reproduce the CMR from Paper I and include the magnitudes 
and colours of the other 10 galaxies observed with the 
ESO/MPG 2.2-m telescope (Section 2.2). We also include 4 galaxies which have 
available colours from PH98. While the two galaxies ESO153-G003 and
ESO194-G021 do not have published velocity dispersion in the literature, and
are excluded from the FP analysis, they have available ($B-R$) colour (Table
1) and are included in the CMR analysis. The three galaxies NGC 3562, NGC 6411
and IC 1211 do not have $R$ magnitudes available in PH98 and are not considered
in the CMR study (see Table 2). 
The residuals of our isolated galaxies from the CMR of galaxies in Coma
cluster show a mean value of 0.05 with $1\sigma$ dispersion of 0.12 (see the
inset in Fig. 4). Fig. 4 confirms our previous result of Paper I 
that isolated galaxies follow the CMR for galaxies in dense environments.

\begin{figure*}
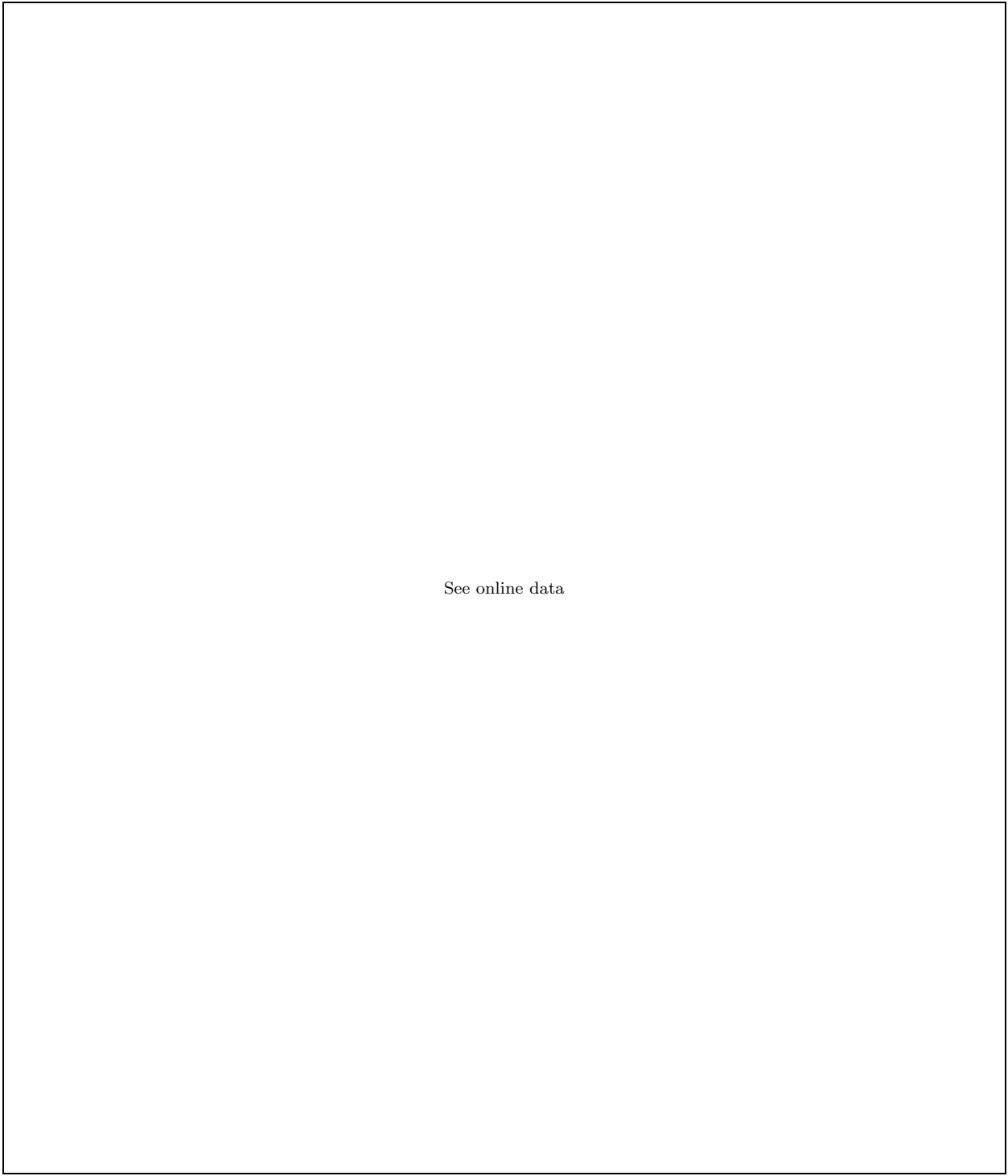

\framebox(470,550)[c]{See online data}
\caption
{
Residual images of galaxies in the $R$-band: NGC 6653, 
NGC 6776, NGC 6799, NGC 6849, NGC 7796, MCG-01-03-018, ESO107-G004, ESO194-G021
and ESO462-G015.
Dust regions can be seen as bright features and extra light as 
dark features. All images have the same size of 188 $\times$ 155 sq. 
arcsec and oriented as North up and East to the left.
}
\label{plot}
\end{figure*}

\begin{figure*}
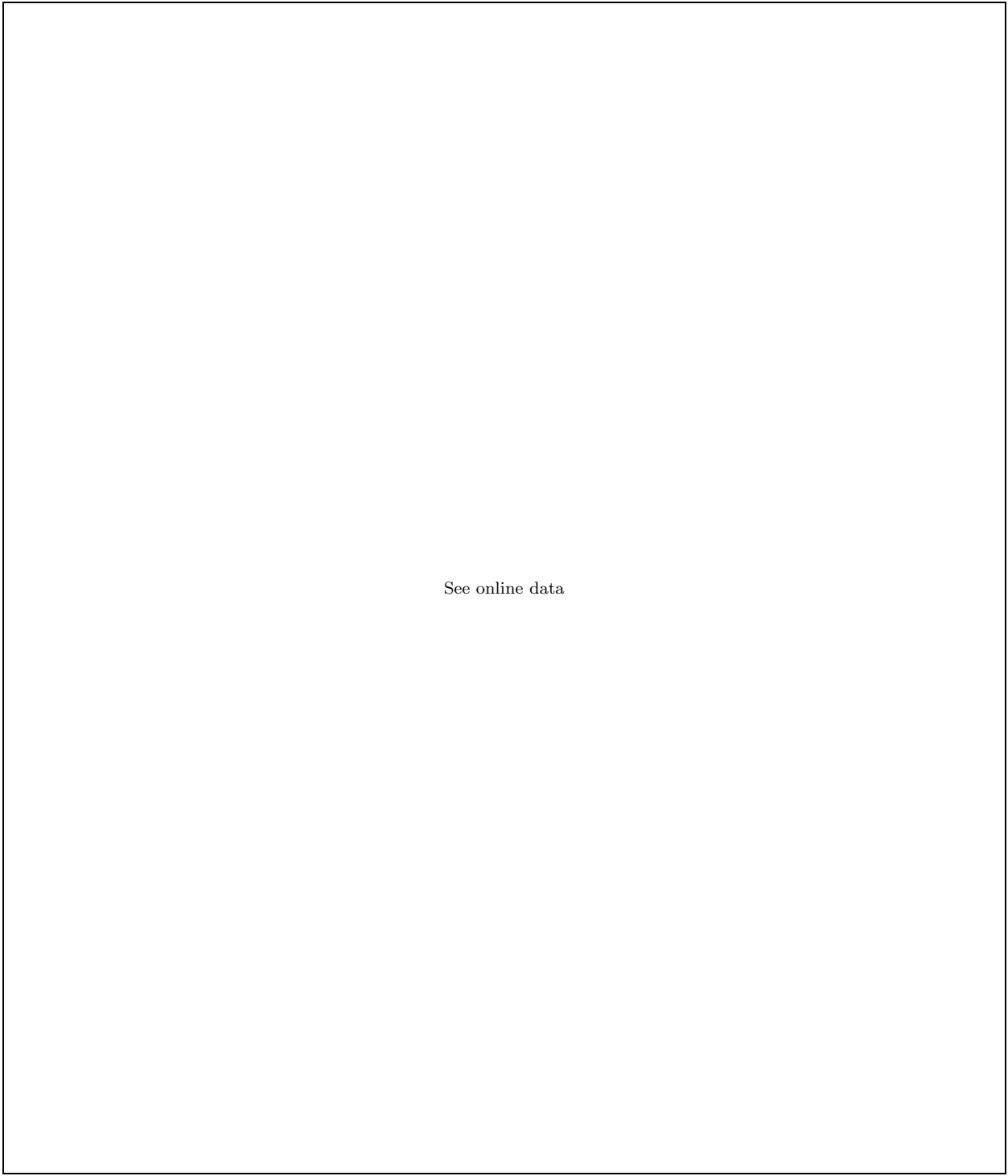

\framebox(470,550)[c]{See online data}
\caption
{
The $R$-band (open circles) surface brightness, ellipticity, position angle 
and 4$^{th}$-order cosine profiles. The upper panel for each galaxy shows
also the $B$-band surface brightness (solid circles). The name of each
galaxy is written above each set of four panels.
}
\label{plot}
\end{figure*}

\subsection{Internal fine structure}

Using the same technique as Paper I, we have modelled each of the 10 galaxies 
observed at ESO/MPG 2.2-m telescope (Sec. 2.2). 
After fitting elliptical isophotes to the galaxy image we subtracted the 
model of each galaxy in the $R$-band to form a residual image and then we 
examined their isophotal shape parameters.   
The internal structure of 9 galaxies (excluding ESO153-G003) is explored 
using their residual images (Fig. 5) and the radial profile of their isophotal
shape parameters (Fig. 6). 

The galaxy ESO153-G003 is saturated in our images which prevents us detecting 
any internal fine structure and it shows no obvious extra tidal light. 
We briefly discuss the remaining 9 galaxies in turn. 
NGC 6653 shows a boxy structure in its outer part as revealed by a negative 
$4^{th}$ cosine term.
NGC 6776 has probable dust within the central region of radius 
$\lesssim 3$ kpc ($8^{''}$). It also reveals 
shell structures and extensive extra tidal light to the West and South.
There is also a tidal tail extending in the southern direction.
In both the $B$ and $R$-bands, our images of NGC 6799 show a smooth elliptical 
profile and no morphological fine structure.
NGC 6849 is disky in the outer part and there is some extra tidal light 
to the North. The $4^{th}$ cosine parameter shows 
negative values in the inner 4 kpc ($10^{''}$) due to the effect of 
a foreground star located near the centre of the galaxy. 
Using multi-colour photometry, Saraiva, Ferrari \& Pastoriza (1999) found that 
the isophotes in the $B$-band showed stronger variations in ellipticity,
position angle and elliptical shape compared to the isophotes in the other 
bands. They suggested that NGC 6849 has traces of dust in the central part.
Comparing the isophotal parameters for NGC 6849 from our $B$ and $R$-band 
imaging, we find no evidence of such differences.
NGC 7796 is a boxy galaxy reaching maximum boxiness in the 
inner region within a radius of 2 kpc ($10^{''}$).
MCG-01-03-018 shows traces of probable dust in central region.
ESO107-G004 has a weak disky structure in the outer part at radii 
greater than 2 kpc ($10^{''}$).
ESO194-G021 shows a disk structure between 1 and 3 kpc ($5^{''}-15^{''}$). 
ESO462-G015 shows a uniform elliptical structure of ellipticity 
$\approx 0.3$ within a radius of 9 kpc ($25^{''}$), but becomes disky at 
larger radii. 
Overall, 7 galaxies (out of 9) appear undisturbed with little or no fine 
structure.


\begin{figure}
\centerline{\psfig{figure=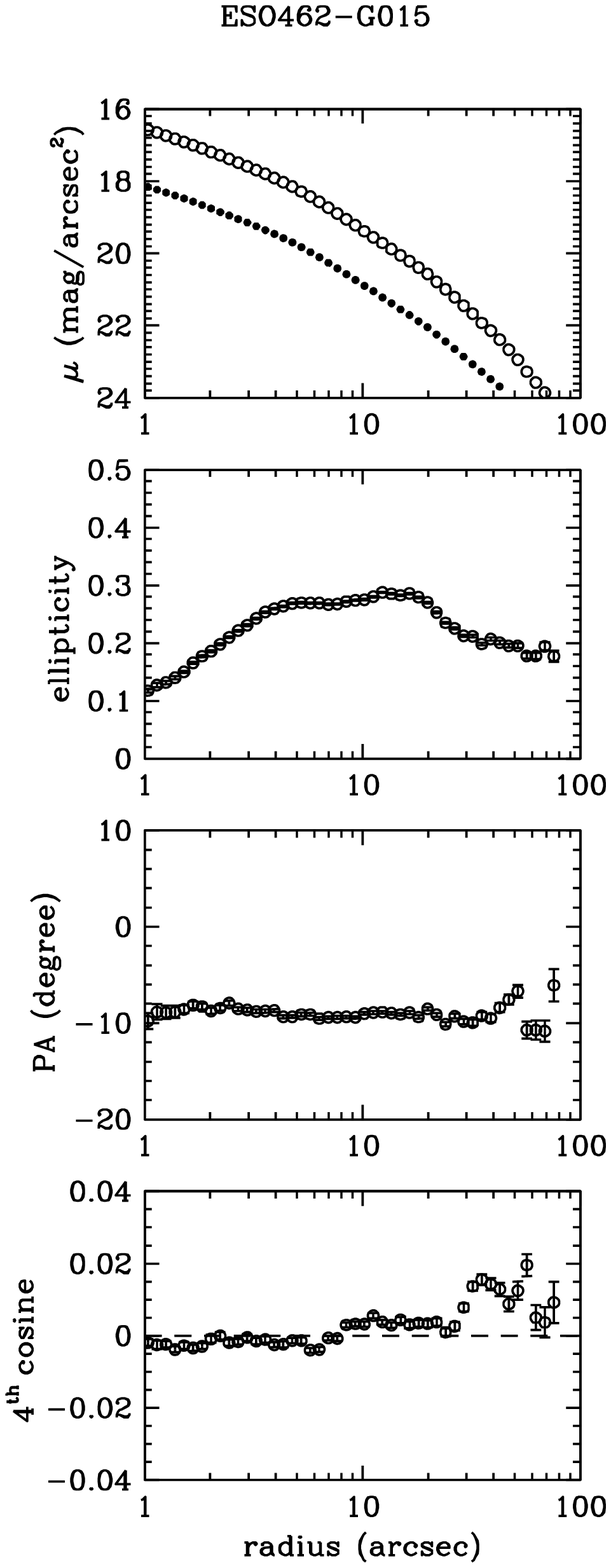,width=0.5\textwidth}}
\contcaption{}
\label{plot}
\end{figure}
 
\subsection{The Fundamental Plane in $\log R_e$, $<\mu>_e$ and 
$\log\sigma$ space}

JFK93 used their imaging observations in the Johnson $B$ and Gunn $r$-bands,
and the central velocity dispersion measurements from Faber et al. (1989) and 
Dressler (1987), for 33 early-type galaxies in the Coma cluster to compute the 
FP tilt and scatter. In the $B$-band, they found a FP of the form: \\
\noindent
$\log R_e = 1.203 \log \sigma + 0.352 <\mu>_e- ~6.642$ ~~~~~~~~~~(1) \\
where $R_e$ values 
are in pc, and the intrinsic scatter is 0.027 dex.  
In Fig. 7, we show the edge-on projection of the FP for our isolated 
early-type 
galaxies using the parameters of JFK93. The solid line is the FP of the 
Coma cluster galaxies (Eq. 1). Considering a typical 
observational error in our data of 0.05, 0.1 and 0.05 in $\log R_e$, 
$<\mu>_B$ and $\log \sigma$ respectively and the intrinsic scatter of Coma 
galaxies, the 1$\sigma$ scatter of the FP is shown in Fig. 7. Our galaxies 
show a similar tilt and scatter as the Coma cluster galaxies, except for the 
four galaxies NGC 2865, NGC 6172, NGC 6776 and NGC 6799 which deviate strongly 
from the FP. The 3 former galaxies are `young' galaxies of age less than 3.2 
Gyrs (TF02; Denicol\'{o} et al. 2005b, D05b),
while NGC 6799 has no published age. We note that such ages 
are luminosity-weighted central ages
based on Lick absorption lines and single stellar population
models to break the age-metallicity degeneracy. Such ages should
not be considered absolute but rather relative ages. Several
other caveats about the application of Lick-style ages to galaxy
populations are discussed in Terlevich \& Forbes (2002). 

Considering the all 23 isolated galaxies in Fig. 7, their residuals
from the FP of the Coma cluster show a mean offset of --0.08 and a
$1\sigma$ dispersion of $\sim 0.15$ (see inset of Fig. 7). About two thirds (15/23)
of our isolated galaxies show negative residuals.

From de la Rosa, de Carvalho \& Zepf (2001), we take data for 12 elliptical 
galaxies in Hickson Compact Groups and 7 galaxies in the field or loose 
groups as a comparison sample. These galaxies cover the same range of $R_e$, 
$<\mu>_e$ and $\sigma$ as our galaxies.
We have excluded galaxies with spectra of S/N $<45$. We also excluded NGC 4552 
which was reported by Caon, Capaccioli \& Rampazzo (1990) 
as a tidally distorted 
elliptical galaxy with an odd luminosity profile. We find that the sample of 
de la Rosa et al. also follows a FP similar to that of the Coma cluster 
ellipticals and the isolated galaxies (Fig. 7). The only exception is 
 NGC 1700 which is also a young galaxy of age $\approx 2$ Gyrs (TF02).

\begin{figure}
\centerline{\psfig{figure=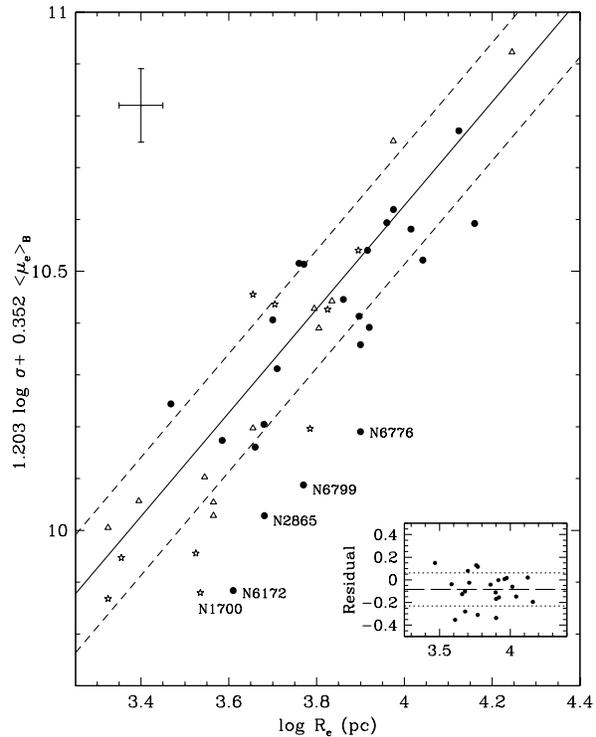,width=0.45\textwidth}}
\caption
{
The edge-on view of the Fundamental Plane for the isolated galaxies in Table 2.
The solid line is the Fundamental Plane of galaxies in Coma cluster from 
J{\o}rgensen, Franx \& Kj{\ae}rgaard (1993). The short-dashed lines represent
the $1\sigma$ dispersion of galaxies in Coma cluster from the same reference. 
Filled circles are our isolated galaxies in Table 2.
Open triangles are galaxies in Hickson Compact Groups and stars are galaxies 
in the field or loose groups (de la Rosa et al. 2001). Galaxies NGC 2865, 
NGC 6172, NGC 6776 and NGC 6799 are the most deviant isolated galaxies from 
the FP and have young ages. The young loose group galaxy  
NGC 1700 also deviates from the FP. A typical error bar is shown.
The inset shows the distribution of the residuals as in Fig. 1. The majority
(15/23) of our isolated galaxies show negative residuals.
}
\label{plot}
\end{figure}


\subsection{ The Fundamental Plane in $\kappa$-space}

By a simple orthogonal coordinate transformation (i.e., a rotation), Bender, 
Burstein \& Faber (1992) introduced a different expression for the FP in terms 
of kappa ($\kappa$) space. The axes of this coordinate system are directly 
proportional to the galaxies physical parameters. 
The edge-on view of the FP is represented by $\kappa_1$ 
and $\kappa_3$, where $\kappa_1$ is a measure of the galaxy mass ($\log M$) 
and $\kappa_3$ is related to the $M/L$ ratio.

 In Fig. 8 we show the distribution of our isolated galaxies in the 
$\kappa_1$-$\kappa_3$ space. They show a similar tilt and dispersion to 
those of Virgo cluster galaxies from Bender et al. (1992). 
The 5 galaxies that show a strong deviation from the FP in Fig. 7, also show a 
tendency to have smaller $\kappa _3$ values i.e. smaller $M/L$ ratio as
expected for their young ages. 
The residuals of the isolated galaxies show a mean value of about --0.01 with
$1\sigma$ dispersion of $\sim 0.11$ (see inset of Fig. 8) which indicates a good
symmetry of our isolated galaxies about the FP of the Coma cluster
in the $\kappa$ space.

\begin{figure}
\centerline{\psfig{figure=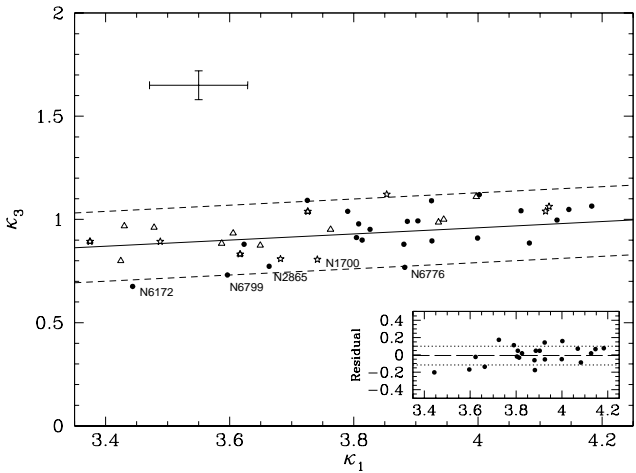,width=0.50\textwidth,angle=-90}}
\caption
{
The edge-on view of the FP of the isolated galaxies in the 
$\kappa_1$-$\kappa_3$ space.
Filled circles are our isolated sample. The solid line is the FP for 
Virgo cluster galaxies from Bender, Burstein \& Faber (1992). 
The short-dashed lines represent the typical observational errors added to 
the $1\sigma$ intrinsic scatter of 0.05 in $\kappa_3$ for Virgo cluster 
galaxies (Nipoti, Londrillo \& Ciotti 2003). Solid circles are the isolated 
galaxies. Open triangles are galaxies in 
Hickson Compact Groups and stars are 
galaxies in the field or loose groups (de la Rosa et al. 2001).
The highly deviant galaxies from the FP in Fig. 7 show a tendency to have 
smaller $\kappa _3$ values than the average. A typical error bar is shown. 
The inset shows the distribution of the residuals as in Fig. 1
}
\label{plot}
\end{figure}

\subsection{The $<\mu_e>_B$-$R_e$ relation}
\begin{figure}
\centerline{\psfig{figure=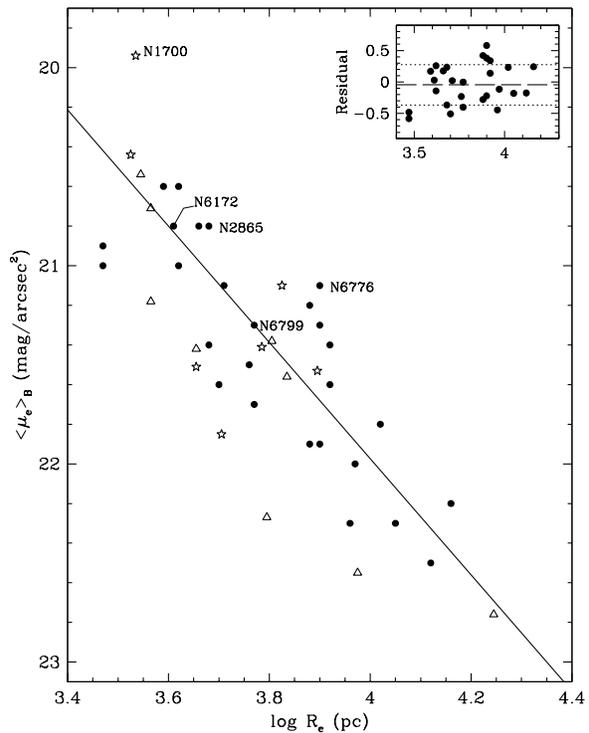,width=0.45\textwidth}}
\caption
{
The projection of the FP in the plane of the mean surface brightness within 
the effective radius $<\mu_e>_B$ and effective radius $R_e$ in the $B$.
The solid line represents the Hamabe \& Kormendy (1987) relation. 
The zero-point of the original relation has been shifted from $V$ to the 
$B$-band assuming typical colours of $B-V = 0.9$.
Symbols as in Fig. 7. The inset shows the distribution of the residuals as in
Fig. 1.  
Our isolated galaxies are generally consistent with the Hamabe \& Kormendy 
relation for luminous galaxies.
}
\label{plot}
\end{figure}

In Fig. 9, we show the projection of the FP in the plane of the mean surface 
brightness within the effective radius $<\mu_e>_B$ and effective radius $R_e$
in the $B$-band. 
The solid line represents the Hamabe \& Kormendy (1987) relation
$\mu_{e}(V) = 2.94 \log($R$_e) + 19.48$
shifted from $V$ to the $B$-band assuming typical colours of $B-V = 0.9$.
We also transformed the surface brightness ($\mu_{e}$) at $R_e$  to 
$<\mu_{e}>$ using the relation of Graham \& Colless (1997).
Our isolated galaxies are generally consistent with the Hamabe \& Kormendy 
relation for luminous galaxies. In the figure we also show the galaxies
in the field and group sample of de la Rosa et al. (2001) which also show 
good consistency with the original relation. 
We note that, while the two galaxies NGC 6172 and NGC 6799 lie on the 
relation, the two isolated galaxies NGC 2865 and NGC 6776 and the group 
galaxy NGC 1700 show significant deviations.

In the inset in Fig. 9 we show the distribution of the residual of our
isolated galaxies from the Hamabe \& Kormendy relation. The mean values of the
residuals show a slight negative offset of 
about --0.05 and $1\sigma$ dispersion of 0.32. 

\section{Discussion}

In this paper we have presented the CMR for 22 isolated early-type galaxies.
We find that the isolated galaxies follow a similar CMR slope and scatter 
to that of galaxies in clusters. Theoretical studies of the CMR indicate that
early-type galaxies formed the bulk of their stars at an early epoch of $z>2$ 
(e.g. Bower, Lucey \& Ellis 1992; Stanford,  Eisenhardt \& Dickinson 1995, 
1998; Ellis et al. 1997; Bower, Kodama \& Terlevich 1998; 
Bernardi et al. 2003c; Tantalo \& Chiosi 2004). None of our isolated galaxies 
show the extremely blue colours seen in the isolated galaxy sample of 
Marcum, Aars \& Fanilli (2004).
On the other hand the isolated galaxy ESO194-G021 shows a very red colour of 
$B-R=1.73 \pm 0.14$ for its luminosity ($M_R=-21.5$).

Considering the morphological investigation of the 9 galaxies in Paper I and 
9 galaxies in the present study, only two galaxies (11 per cent), 
NGC 2865 and NGC 6776, show obvious shell structures. 
This is less than the frequency found by Malin \& Carter (1983), Seitzer \& 
Schweizer (1990) and Reduzzi, Longhetti \& Rampazzo (1996) who quoted 
higher fractions of 17, $>$50 and 16.4 per cent respectively for shell 
galaxies in low-density environments. On the other hand, 
we find more shells than 
Marcum et al. (2004) who detected no shells in any of their 8 
isolated early-type galaxies.
If evidence of a past merger includes shells, dust, plumes, disky and boxy 
structures, then we detected mergers in 60 per cent (11/18) of our isolated 
galaxies. This is a higher fraction than the 44 per cent quoted by 
Reduzzi et al. (1996) for their sample of 61 isolated early-type galaxies. 
Also, a recent study by Michard \& Prugniel (2004) found the  frequency 
of galaxies with perturbed morphologies in poor group environments
to be $\approx 35$ per cent compared to only $\approx 19$ per cent in the 
Virgo cluster. 
About 28 per cent (5/18) of our galaxies contain dust which is 
comparable to the 24.6 per cent of dusty galaxies quoted by 
Reduzzi et al. (1996).
None of our 18 galaxies show irregular structure (radial changes between boxy 
and disky structures within the galaxy) compared to the 50 per cent found by
 Zepf \& Whitmore (1993) for ellipticals in Hickson Compact Groups.
We speculate that the high density of Hickson Compact Groups and 
the resultant high interaction rate gives rise to irregular isophotes whereas 
in isolated environments such interactions do not occur.

In Figures 7 and 8 we have compared the FP for our isolated early-type 
galaxy sample with that for galaxies in higher density environments.  
We find that galaxies in a wide range of environments 
are consistent with the same FP of similar tilt and scatter. 
In Fig. 7, the four galaxies NGC 2865, NGC 6172, NGC 6776 and NGC 6799 of 
our isolated sample and the group galaxy NGC 1700 of de la Rosa et al. 
(2001) deviate strongly from the main trend of the FP. 
In Fig. 8, the same five galaxies show a tendency to have smaller $\kappa_3$ 
in the direction of lower $M/L$ ratios which can be explained by the young age 
of their stellar population.

We note that the isolated galaxies NGC 2865 and NGC 6776 and the group 
galaxy NGC 1700 also deviate from the $<\mu_e>_B$-$R_e$ relation 
(Fig. 9). This indicates either a relatively large effective radius or high 
surface brightness.
The isolated galaxies NGC 6172 and NGC 6799 lie on the Kormendy relation. 
The deviation of these two galaxies from the FP can be accounted for by their 
small velocity dispersions which are about 75 per cent of the expected 
values for their luminosities (Forbes \& Ponman 1999). 

The relative mean age of our galaxy sample can be estimated using the values
quoted in TF02 and D05b.  
In their catalogues, the authors used H$\beta$ and [MgFe] absorption 
line indices to break the age/metalicity degeneracy. 
These ages reflect the young stars in the central regions which were presumably
formed during the last gaseous merger event. Therefore, the measured 
ages of these stellar populations give an estimate of the time elapsed 
since the last gaseous merger.
Twelve galaxies (out of 23) of our sample have available published ages.
The mean age of these galaxies is 4.6 $\pm 1.4$ Gyr, which is approximately
similar to the mean age quoted by Proctor et al. (2004) for galaxies in 
small groups and field ($5.9 \pm 0.7$ Gyr) but younger than galaxies in 
cluster environments ($> 8.5 \pm 0.7$ Gyr).

In Fig. 7, we note that the most deviant galaxies from the FP are the 4 
youngest galaxies of our sample with ages $\lesssim 3$ Gyr. 
NGC 2865 has a relatively 
blue colour of $B-R=1.3$ (Paper I) and reveals many fine structures such as 
shells, tidal light, dust and a kinematically distinct core 
(Paper I; Hau, Carter \& Balcells 1999) which implies a past 
merger involving at least one gas-rich progenitor. Hau et al. (1999) quoted an 
age estimate of $\sim 1.1$ Gyr since the last merger which is
comparable to the stellar population age of $< 1.5$ Gyr estimated by TF02.
NGC 6776 has a tidal tail, shells and extensive extra tidal light. 
The tidal tail suggests at least one of the progenitors was a disk galaxy. 
In a detailed photometric and spectroscopic study of NGC 6776, Sansom, Reid \&
Boisson  (1988) measured a rapid rotation of $\approx 100$ km/s and a velocity 
dispersion of $\approx 200$ km/s . 
They detected no dust or young stars which led them to conclude that 
the merger occurred $\geqslant 1$ Gyr ago. TF02 measured an age of 3.2 
Gyr for its stellar population.

While the residual image of NGC 6172 (Paper I) did not reveal any obvious 
features, the unsharp-masking and colour map by Colbert, Mulchaey \& 
Zabludoff (2001) 
revealed a weak shell and some dust near the centre of the galaxy. 
Its global blue colour ($B-R=1.3$; Paper I) and the young 
age of 1.6 Gyr (D05b) suggest a recent starburst induced by a merger.
NGC 6799 has a large dust lane along its eastern edge (Colbert et al. 2001). 
The residuals from the FP, using the method of PS96, is $-0.37$ which 
suggests the presence of a young stellar population of age $\lesssim$ 1.5 Gyr
(Forbes, Ponman \& Brown 1998). This evidence supports the 
suggestion of a merger past for NGC 6799.

Despite the young age of 2.7 and 5.4 Gyr for NGC 1045 and ESO462-G015 
respectively (D05b), and the tidal tail that was 
detected in NGC 1045 (Paper I) indicating a recent merger, they both lie 
within 1$\sigma$ of the FP.
The two galaxies NGC 6849 and MCG-01-03-018, have very old ages (TF02; D05b). 
The old age for their central stellar populations and the absence of fine 
structures suggests that there has not been a gaseous 
merger in the recent past for these two galaxies.
The galaxy NGC 1132 reveals an extended group-like X-ray structure. 
This lead Mulchaey \& Zabludoff (1999) to speculate that this galaxy was 
a remnant of a merged group of galaxies (a fossil). 
The absence of fine structure and an old central stellar population (D05b) 
suggests that it formed at early epochs and has not accreted any gas-rich 
galaxies recently.

Numerical simulations show that dissipational mergers between two disk, 
star-forming and gas-rich galaxies can produce non-homologous galaxies 
reproducing the observed tilt and scatter of the FP (Bekki 1998). 
Dissipationless mergers can also reproduce the FP (Hjorth \& Madsen 1995; 
Capelato et al. 1995; Levine 1997; Dantas et al. 2003; Nipoti et al. 2003). 
However, as Nipoti et al. (2003) point out, dissipationless 
merging has difficulty explaining other scaling relations such as the 
colour-magnitude and black hole mass-$\sigma$ relations. 
Unlike major mergers, neither accretion of smaller galaxies or a simple 
monolithic collapse are plausible scenarios to reproduce the FP 
(Dantas et al. 2003; Nipoti et al. 2003), however more realistic collapse 
conditions need to be explored.

\section{Summary and conclusion}

In this paper we show that isolated early-type galaxies 
follow a colour-magnitude relation of similar slope and scatter to that of 
early-type galaxies in the dense environments of the Coma cluster. 
This would suggest that early-type galaxies formed the bulk of their stars 
at $z>2$ (10.3 Gyrs ago in the $\Lambda$CDM cosmology).
The scatter in this relation is explained as a result of a secondary 
starburst, resulting from a gaseous merger at a later epoch. 

Two galaxies (11 per cent) of our isolated galaxies sample reveal shells.
On the other hand, about 60 per cent (11/18) of our isolated galaxies revealed 
evidence of past mergers such as shells, dust, plumes, disky and boxy 
structures. Four of these eleven galaxies are the youngest members of our 
sample with ages $\lesssim 3$ Gyrs.

We also present the Fundamental Plane for our sample of isolated early-type 
galaxies. It shows a similar tilt and scatter to that for galaxies in denser 
environments. 
However some galaxies deviate from the relation due to the young age of their 
stellar populations or lower velocity dispersions.
These young galaxies also show tendency to have smaller mass-to-light ratios 
in $\kappa$-space.
The distribution of these galaxies in the $<\mu_e>_B$-$R_e$ (Kormendy) 
projection is consistent with this interpretation. 

Galaxies in such isolated environments show a relatively young mean age which  
is similar to that of galaxies in the field and small groups, but is younger 
than that for galaxies in clusters and Hickson Compact Groups. 
This result is qualitatively consistent with the expectations of the 
hierarchical galaxy formation.

From these results we conclude that relatively recent mergers offer a 
plausible explanation for the observed photometric and kinematic properties of 
{\it some} isolated early-type galaxies. 
For other isolated galaxies, which show neither fine structure or young 
stellar populations, we speculate that they formed at early epochs and evolved 
passively thereafter.

\

\noindent{\bf ACKNOWLEDGEMENTS}
\

\noindent
Thanks to Paul Goudfrooij and Vera Kozhurina-Platais for supplying us with 
the reduced images observed with the ESO/MPG telescope. We would also like to 
thank Ale Terlevich for his help with the observations.
We wish to thank Dr. S. Brough for her useful comments in the final 
revision of this paper.

\

\noindent{\bf REFERENCES}

\noindent
Bekki K., 1998, ApJ, 496, 713 \\
Bender R., Burstein D., Faber S. M., 1992, ApJ, 399, 462 \\
Bernardi M. et al., 2003a, AJ, 125, 1849\\
Bernardi M. et al., 2003b, AJ, 125, 1866\\
Bernardi M. et al., 2003c, AJ, 125, 1882\\
Bower R. G., Lucey J. R., Ellis R. S., 1992, MNRAS, 254, 601\\
Bower R. G., Kodama T., Terlevich A., 1998, MNRAS, 299, 1193\\
Caon N., Capaccioli M., Rampazzo R., 1990, A\&AS, 86, 429 \\
Capelato H. V., de Carvalho R. R., Carlberg R. G., 1995, ApJ, 451, 525 \\
Colbert J.W., Mulchaey J.S., Zabludoff A.I., 2001, AJ, 121, 808 \\
Dantas C. C., Capelato H. V., Ribeiro A. L. B., de Carvalho R. R., 2003, 
    MNRAS, 340, 398 \\
de Carvalho R. R., Djorgovski S., 1992, ApJ, 389, 49\\
de la Rosa I. G., de Carvalho R. R., Zepf S. E., 2001, AJ, 122, 93\\
Denicol\'o G, Terlevich R, Terlevich E., Forbes D. A., Terlevich A., 
      Carrasco L., 2005a, MNRAS, 356, 1440 (D05a) \\
Denicol\'o G, Terlevich R, Terlevich E., Forbes D.A., Terlevich A., 
       2005b, MNRAS, 358, 813 (D05b)\\
Djorgovski S., Davis M., 1987, ApJ, 313, 59 \\
Djorgovski S., de Carvalho R., Han M. S. , 1988, in Pritchet C. J., 
van der Bergh S., eds, ASP Conf. Ser. Vol. 4, The Extragalactic Distance Scale.
 Astron. Soc. Pac., San Francisco, p. 329 \\
Dressler A., 1987, ApJ, 317, 1 \\
Ellis R.S., Smail I., Dressler A., Couch W.J., Oemler A., Butcher H., 
    Sharples, R.M., 1997, ApJ, 483, 582\\
Evstigneeva E.A., Reshetnikov V.P., Sotnikova N. Ya., 2002, A\&A, 381, 6\\
Faber S.M., Wegner G., Burstein D., Davies R.L., Dressler A., 
       Lynden-Bell D., Terlevich R.J., 1989, ApJS, 69, 763 \\
Forbes D.A., Ponman T., 1999, MNRAS, 309, 623\\
Forbes D.A., Ponman T.J., Brown R.J.N., 1998, ApJ, 508, 43 \\
Gladders M.D., Lopez-Cruz O., Yee H.K.C., Kodama T., 1998, ApJ, 501, 571 \\
Graham A., Colless M., 1997, MNRAS, 287, 221\\
Hamabe M., Kormendy J., 1987, in de Zeeuw T., ed., Proc. IAU Symp. 127, 
Structure and Dynamics of Elliptical Galaxies. Reidel, Dordrecht, p. 379\\
Hau G.K.T., Carter D., Balcells M., 1999, MNRAS, 306, 437\\
Hau G.K.T., Forbes D.A., 2005, in preparation (HF05) \\
Hjorth J., Madsen J., 1995, ApJ, 445, 55 \\
J{\o}rgensen I., Franx M., Kj{\ae}rgaard P., 1993, ApJ, 411, 34 (JFK93) \\
J{\o}rgensen I., Franx M., Kj{\ae}rgaard P., 1995, MNRAS, 276, 1341 (JFK95)\\
Kauffmann G., White S.D. M., Heckman T.M., Ménard B., Brinchmann J., 
Charlot S., Tremonti C., Brinkmann J., 2004, MNRAS, 353, 713 \\
Koprolin W., Zeilinger W.W., 2000, A\&AS, 145, 71 \\
Landolt A. U., 1992, AJ, 104, 340 \\
Levine S., 1997, The Nature of Elliptical Galaxies; 2nd Stromlo Symposium. 
                 ASP Conference Series; Vol. 116; 1997; ed. 
                 Arnaboldi M., Da Costa G.S. and Saha P., p.166\\
Malin D.F., Carter D., 1983, ApJ, 274, 534 \\
Marcum P. M., Aars C. E., Fanelli M. N., 2004, AJ, 127, 3213 \\
Michard R., Prugniel P., 2004, A\&A, 423, 833\\
Mulchaey J.S., Zabludoff A.I., 1999, ApJ, 514, 133\\
Nipoti C., Londrillo P., Ciotti L., 2003, MNRAS, 342, 501 \\
Proctor R. N., Forbes D. A., Hau G. K. T., Beasley M. A., De Silva G. M., 
       Contreras R., Terlevich A. I., 2004, MNRAS, 349, 1381\\
Prugniel Ph., Heraudeau Ph., 1998, A\&ASS, 128, 299 (PH98) \\
Prugniel Ph., Simien F., 1996, A\&A, 309, 749 (PS96) \\
Reda F.M., Forbes D.A., Beasley M., O'Sullivan E.J., Goudfrooij P., 2004, 
     MNRAS, 354, 851 (Paper I) \\
Reduzzi L., Longhetti M., Rampazzo R., 1996, MNRAS, 282, 149 \\
Sansom A. E., Reid I. Neill, Boisson C., 1988, MNRAS, 234, 247\\
Saraiva M.F., Ferrari F., Pastoriza M.G., 1999, A\&A 350, 399 \\
Schlegel D.J., Finkbeiner D.P., Davis M., 1998, ApJ, 500, 525\\
Seitzer P., Schweizer F., 1990, Dynamics and Interactions of Galaxies, 
     ed. R. Weilen (New York: Springer), 270\\
Stanford S.A., Eisenhardt P.R.M., Dickinson M.E., 1995, ApJ, 450, 512\\
Stanford S.A., Eisenhardt P.R.M., Dickinson M.E., 1998, ApJ, 492, 461\\
Tantalo R., Chiosi C., 2004, MNRAS, 353, 405\\
Terlevich A.I., Forbes D.A., 2002, MNRAS, 330, 547 (TF02) \\
Trujillo I., Burkert A., Bell E.F., 2004, ApJ, 600, 39 \\
Zepf S.E., Whitmore B.C., 1993, ApJ, 418, 72\\
\end{document}